# Partial Identification of Personalized Treatment Response with Trial-reported Analyses of Binary Subgroups

Sheyu Li[a], Valentyn Litvin[b], and Charles F. Manski[c]

September 2022

Abstract

Medical journals have adhered to a reporting practice that seriously limits the usefulness of published trial findings. Medical decision makers commonly observe many patient covariates and seek to use this information to personalize treatment choices. Yet standard summaries of trial findings only partition subjects into broad subgroups, typically into binary categories. Given this reporting practice, we study the problem of inference on *long* mean treatment outcomes E[y(t)|x], where t is a treatment, y(t) is a treatment outcome, and the covariate vector x has length K, each component being a binary variable. The available data are estimates of $\{E[y(t)|x_k = 0], E[y(t)|x_k = 1], P(x_k)\}$, k = 1, . . . , K reported in journal articles. We show that reported trial findings partially identify {E[y(t)|x], P(x)}. Illustrative computations demonstrate that the summaries of trial findings in journal articles may imply only wide bounds on long mean outcomes. One can realistically tighten inferences if one can combine reported trial findings with credible assumptions having identifying power, such as *bounded-variation* assumptions.

[a] Department of Endocrinology and Metabolism, MAGIC China Centre, Cochrane China Centre, Chinese Evidence-based Medicine Centre, West China Hospital, Sichuan University

[b] Department of Economics, Northwestern University

[c] Department of Economics and Institute for Policy Research, Northwestern University

We thank Gordon Guyatt for inspiring this study idea. Ivan Canay, Francesca Molinari, Qingyang Shi, and Joerg Stoye provided helpful comments.

## 1. Introduction

Our concern with the inferential problem studied in this paper stems from difficulties we have encountered when seeking to interpret summaries of trial findings reported in medical journals. Modern patient care aims to base treatment choice on the findings of randomized trials that compare alternative treatments. However, journals have adhered to a reporting practice that seriously limits the usefulness of published findings. Medical decision makers commonly observe many patient covariates and seek to use this information to personalize treatment choice. Yet standard summaries of trial findings only partition subjects into broad subgroups, typically into binary categories.

*Illustration*: Zinman *et al.* (2015) studied the effects of the drug empagliflozin on cardiovascular morbidity and mortality in patients with type 2 diabetes at high cardiovascular risk. The investigators randomly assigned subjects to receive empagliflozin or placebo once daily and compared the between-group incidence of the primary composite outcome, which included cardiovascular death from cardiovascular causes, nonfatal myocardial infarction, and nonfatal stroke. Like most other trials, the authors followed the reporting guideline (Schulz, *et al.,* 2010) and reported multiple subgroup analyses defined by categorical covariates, most of which were binary including sex, age, and high or low glycated hemoglobin (HbA1c).

In practice, a diabetologist knows patient values for most or all of these covariates. Thus, learning the probability of the outcome for a patient with a specified combination of covariates (age, sex, HbA1c, etc.) would facilitate decision-making. However, the journal article only reported findings on the frequency of the outcome in each covariate category separately. Note that the study population was not geographically comprehensive, which limits clinical interpretation in particular countries (Zhou *et al*., 2021).

Constraints on article length set by journals make it unrealistic to expect more than marginal additions to the summaries of trial findings now presented in main texts. Concern with type I error (false positives) also restricts reporting multiple subgroup analysis (Wang *et al.*, 2007). Individual-level patient data would be very helpful but its sharing is rare in practice.

Given the standard reporting practice, this paper studies the problem of inference on mean treatment outcomes E[y(t)|x]. Here t is a treatment administered in a randomized trial and y(t) is a bounded real-valued treatment outcome. The covariate vector $x = (x_1, x_2, \ldots, x_K)$ has length K, each component being a binary categorical variable. The available data are estimates of $\{E[y(t)|x_k = 0], E[y(t)|x_k = 1], P(x_k)\}$, k = 1, . . . , K reported in journal articles. Adapting the regression terminology of Goldberger (1991) and Cross and Manski (2002), we refer to E[y(t)|x] and $E[y(t)|x_k]$ as *long* and *short* mean outcomes.

The present inferential problem is reminiscent of but differs from one analyzed in Cross and Manski (2002) and Manski (2018). There the aim was similarly to learn a long regression E(y|x). What differed was the available data. The data were estimates of short conditional distributions $P(y|x_1, x_2, \ldots x_J)$ and $P(x_{J+1}, x_{J+2}, \ldots, x_K|x_1, x_2, \ldots x_J)$ for some J such that $1 \leq J < K$. Inference on long regressions using estimates of these short distributions has commonly been called *ecological inference*.

Here, as in ecological inference, the basic problem is identification. Section 2 shows that knowledge of $\{E[y(t)|x_k = 1], E[y(t)|x_k = 0], P(x_k)\}$, k = 1, . . . , K partially identifies {E[y(t)|x], P(x)}. The identification region is the set of solutions to equations that relate short to long mean outcomes and inequalities that bound the possible values of the unknown quantities.

Our abstract identification analysis is simple, but computation of identification regions poses challenges. We ise a tractable numerical method to compute sharp bounds on long mean outcomes. We use trial findings in Zinman *et al.* (2015) to illustrate. Computation in this illustrative case and others demonstrates that the summaries of trial findings in journal articles may imply only very wide, indeed often uninformative, bounds on long mean outcomes.

One might anticipate that observed properties common to short mean outcomes imply corresponding properties of long mean outcomes. To the extent that this holds, the summary of trial findings reported in journal articles may be useful. However, we find that important properties cannot be extrapolated from short to long mean outcomes.

For example, suppose that a reported summary of trial findings reveals that, for a given $\xi \in \{0, 1\}^K$ and real number V, $E[y(t)|x_k = \xi_k] > V$ for all $k = 1, \ldots, K$. One might then expect that $E[y(t)|x = \xi] > V$, but this does not follow. The value of the long mean outcome may be substantially less than V.

One can realistically tighten long inferences if one can combine reported trial findings with credible assumptions having identifying power. Section 3 poses and studies *bounded-variation* assumptions. Such assumptions have previously been used in ecological inference. We again use data from Zinman *et al.* (2015) to illustrate.

While the core concern of this paper is identification, it is also desirable to measure sampling imprecision. The prevalent approach to sample inference on partially-identified parameters has been to compute confidence sets with desirable asymptotic properties. But the methods developed to date do not fit the present setting. Section 4 explains and suggests an alternative way to measure imprecision.

The reporting issues examined in this paper could be mitigated if trial investigators were to make public the data they collect on subject-specific covariates and outcomes, with appropriate steps taken to ensure privacy. Section 5 recommends that this occur.

Throughout the paper, we use medical illustrations of the inferential problem because reporting short trial findings is so prevalent in medical research. However, the problem arises outside of medicine as well. It has, for example, become common for empirical microeconomists to emulate medical research in the design, analysis, and reporting of randomized trials. See American Economic Association (2022a).

## 2. Identification of Long Mean Outcomes Using Short Trial Findings

Henceforth, let y(t) be a bounded real outcome under treatment t, whose range is normalized to take values in the interval [0, 1]. Let $x = (x_1, x_2, \ldots, x_K)$ be a K-variate vector of binary covariates, each taking the value 0 or 1. The objective is to learn $E[y(t)|x = \xi]$, $\xi \in \{0, 1\}^K$. Summaries of trial findings in research articles commonly provide estimates of the K short mean outcomes $E[y(t)|x_k = 0]$ and $E[y(t)|x_k = 1]$, $k = 1, \ldots, K$. Articles also commonly report the fraction of the study sample having each binary covariate value. These fractions provide estimates of $P(x_k)$, $k = 1, \ldots, K$.

2.1. Identification Analysis

The basic identification problem is to determine what can be deduced about $E[y(t)|x]$ and $P(x)$, given knowledge of $[E[y(t)|x_k = 0], E[y(t)|x_k = 1], P(x_k)]$, $k = 1, \ldots, K$. In general, $E[y(t)|x]$ and $P(x)$ are partially rather than point identified. To focus on identification, we suppose that reported trial estimates are accurate.

We begin with inequalities satisfied by $E[y(t)|x]$ and $P(x)$. A covariate vector may be relevant to treatment only if it occurs with positive probability. Hence, $E[y(t)|x]$ satisfy the inequalities

$$\text{(1a)} \quad 0 \leq E[y(t)|x = \xi] \leq 1, \quad \xi \in \{0, 1\}^K,$$

and we suppose that

$$\text{(1b)} \quad 0 < P(x = \xi), \quad \xi \in \{0, 1\}^K.$$

Next consider the relationship between short and long mean treatment outcomes. Let $x_{-k}$ be the $(K-1)$-dimensional sub-vector of x that excludes $x_k$. The Law of Iterated Expectations and Bayes Theorem give 2K equations relating short to long mean outcomes:

$$\text{(2a)} \quad E[y(t)|x_k = 0]\cdot P(x_k = 0) = \sum_{\xi_{-k} \in \{0, 1\}^{K-1}} E[y(t)|x_k = 0, x_{-k} = \xi_{-k}]P(x_k = 0, x_{-k} = \xi_{-k}),$$

(2b) $E[y(t)|x_k = 1]\cdot P(x_k = 1) = \sum_{\xi_{-k} \in \{0, 1\}^{K-1}} E[y(t)|x_k = 1, x_{-k} = \xi_{-k}]P(x_k = 1, x_{-k} = \xi_{-k})$.

The Law of Total Probability gives K equations relating marginal to joint probabilities of covariate values and one more from the fact that joint probabilities sum to one:

(3a) $P(x_k = 1) = \sum_{\xi_{-k} \in \{0, 1\}^{K-1}} P(x_k = 1, x_{-k} = \xi_{-k})$,

(3b) $1 = \sum_{\Xi \in \{0, 1\}^{K}} P(x = \xi)$.

Reported summaries of trial findings reveal the left-hand sides of equations (2)−(3). The quantities $E[y(t)|x]$ and $P(x)$ on the right-hand sides are not reported in journal articles. Inequalities (1) and equations (2)-(3) express all of the available information about these quantities. Thus, the identification region for $\{E[y(t)|x], P(x)\}$ is the set of values that solve (1)−(3).

Observe that Equations (2)−(3) comprise 3K+1 equations in $2^{K+1}$ unknowns. Equations (2) connect the problems of identification of $E[y(t)|x]$ and $P(x)$. If the marginal covariate probabilities $[P(x_k), k = 1. , . . , K]$ were observed but not the short mean outcomes $\{E[y(t)|x_k], k = 1, . . . , K\}$, no conclusions could be drawn about $E[y(t)|x]$. Inference on $P(x)$ using (1b) and (3) would be the classical problem of identification of a multivariate distribution given knowledge of its marginals, studied in the literature on Fréchet bounds.

Analysis of identification of long average treatment effects (ATEs) and relative risks is a simple extension of the above. Whereas we stated inequalities (1a) and equations (2) for a specified treatment t, analogous conditions hold for any other treatment t′. Inequalities (1b) and equations (3) hold as stated when considering t′. Thus, the feasible value of $E[y(t)|x] – E[y(t')|x]$, $E[y(t)|x]/E[y(t')|x]$, and $P(x)$ are those that solve (1)−(3) and the analogous versions of (1b) and (2) applied to t′.

## 2.2. Computation of Sharp Bounds on Long Mean Treatment Outcomes

Although the above identification analysis is simple, computation of identification regions is complex. Equations (2) are nonlinear, with multiplicative terms $E[y(t)|x = \xi]\cdot P(x = \xi)$, $\xi \in \{0, 1\}^K$. It follows that the identification region for $\{E[y(t)|x], P(x)\}$ is not convex. To see this, suppose one finds that specific values $\{E_0[y(t)|x], P_0(x)\}$ and $\{E_1[y(t)|x], P_1(x)\}$ are feasible. If the identification region were convex, the convex combination $c\{E_1[y(t)|x], P_1(x)\} + (1 - c)\{E_0[y(t)|x], P_0(x)\}$ would be feasible for all $c \in (0, 1)$. This long vector satisfies (3) but not (2). The multiplicative terms $E[y(t)|x]\cdot P(x)$ on the r.h.s. of (2) imply sub-additivity, as $c^2 + (1 - c)^2 < 1$.

Rather than take on the formidable problem of computing the joint identification region for $\{E[y(t)|x = \xi], P(x) = \xi\}$ for all values of $(t, \xi)$, we specify a value of $(t, \xi)$ and address the still challenging but more tractable problem of computing sharp lower and upper bounds for the scalar $E[y(t)|x = \xi]$. This task entails minimizing and maximizing $E[y(t)|x = \xi]$ subject to the constraints imposed by (1)−(3). The identification region for $E[y(t)|x = \xi]$ necessarily is a subset of the interval connecting the sharp bounds. Appendix 1 shows that the region for $E[y(t)|x = \xi]$ is the entire interval when the outcome y(t) is binary. Appendix 2 describes our computational approach.

## 2.3. Illustrative Application

To illustrate, we use findings in Zinman *et al.* (2015). The computations that we perform abstract from the difference between the estimates in the article and population probabilities. Sampling imprecision of the estimates is a relatively minor concern in this illustration because the identification problem is severe and because the trial size was large, with 4687 subjects receiving empagliflozin and 2333 receiving placebo.

Let treatment t be empagliflozin, labelled t = e. The primary outcome yI is binary, so $I(e)|x] = P[y(e) = 1|x]$. Let K = 3. Let $x_1 = 0$ or 1 if age < 65 or ≥ 65 years, labeled Y or O. Let $x_2 = 0$ or 1 if sex is male or female, labeled M or F. Let $x_3 = 0$ or 1 if glycated hemoglobin is < 8.5% or ≥ 8.5%, labelled L or H. For example, x = (0,1,0) labelled YFL, represents females of age < 65 years with glycated hemoglobin < 8.5%.

Data in Table S7 of the published article enables calculation of estimates of the probabilities that each covariate takes the value 0 or 1. Each estimate is the frequency of the covariate value in the group of subjects who receive empagliflozin. The estimates for $x_k = 1$ are

$$P(x_1 = 1) = 0.446, \quad P(x_2 = 1) = 0.288, \quad P(x_3 = 1) = 0.315.$$

The data enable calculation of estimates of short mean treatment outcomes, these being the frequencies with which the primary outcome occurs within the group of subjects who receive empagliflozin and have a specified covariate value. The estimates are

$$P[y(e) = 1|x_1 = 0] = 0.097, \quad P[y(e) = 1|x_1 = 1] = 0.114,$$

$$P[y(e) = 1|x_2 = 0] = 0.110, \quad P[y(e) = 1|x_2 = 1] = 0.091,$$

$$P[y(e) = 1|x_3 = 0] = 0.100, \quad P[y(e) = 1|x_3= 1] = 0.114.$$

These short mean outcomes all lie within the narrow range from 0.091 to 0.114. A reader of Zinman *et al.* (2015) may think it reasonable to extrapolate that long mean outcomes must lie approximately within this range as well. But this conclusion does not follow from the available data. Given only the data and no additional assumptions, we find that the sharp bounds on long mean outcomes differ almost imperceptibly from the trivial bounds [0, 1]. The results are in Table 1A.

Table 1A: Sharp Bounds on Long Mean Outcomes with Treatment by Empagliflozin

| **x label** | **P[y(e) = 1\|x] lower bound** | **P[y(e) = 1\|x] upper bound** |
|---|---|---|
| YML | 0.000 | 1.000 |
| OML | 0.000 | 1.000 |
| YFL | 0.000 | 1.000 |
| OFL | 0.000 | 1.000 |
| YMH | 0.001 | 0.999 |
| OMH | 0.000 | 1.000 |
| YFH | 0.000 | 1.000 |
| OFH | 0.000 | 1.000 |

We have computed analogous bounds on long mean outcomes with treatment by placebo; denoted t = p. In this case, the estimates for $x_k = 1$ are

$$P(x_1 = 1) = 0.444, \quad P(x_2 = 1) = 0.280, \quad P(x_3 = 1) = 0.311.$$

The estimates of short mean outcomes are

$$P[y(p) = 1|x_1 = 0] = 0.093, \quad P[y(p) = 1|x_1 = 1] = 0.155,$$

$$P[y(p) = 1|x_2 = 0] = 0.126, \quad P[y(p) = 1|x_2 = 1] = 0.107,$$

$$P[y(p) = 1|x_3 = 0] = 0.130, \quad P[y(p) = 1|x_3= 1] = 0.101.$$

The sharp bounds on long mean outcomes are again extremely wide. The solver results are in Table 1B.

Table 1B: Sharp Bounds on Long Mean Outcomes with Treatment by Placebo

| **x label** | **P[y(p) = 1\|x] lower bound** | **P[y(p) = 1\|x] upper bound** |
|---|---|---|
| YML | 0.000 | 1.000 |
| OML | 0.020 | 1.000 |
| YFL | 0.000 | 1.000 |
| OFL | 0.001 | 1.000 |
| YMH | 0.001 | 1.000 |
| OMH | 0.000 | 1.000 |
| YFH | 0.000 | 0.999 |
| OFH | 0.000 | 1.000 |

## 3. Identification with Bounded-Variation Assumptions

We have performed computations with a spectrum of short trial findings and found that the implied bounds on long mean outcomes are typically wide. Sometimes they are so wide as to be essentially uninformative, as are those presented above using the findings in Zinman *et al.* (2015). We therefore strongly caution against loose extrapolation from short findings to long mean outcomes. Nevertheless, we do not conclude that reported short findings are useless to medical decision making.

One can tighten long inferences in a realistic way if one can combine reported trial findings with credible assumptions that have identifying power. Such assumptions may take many forms. We discuss *bounded-variation* assumptions. These have previously been used to add identifying power in ecological inference (Manski, 2018; Manski, Tambur, and Gmeiner, 2019) and other settings with partial identification of treatment effects (Manski and Pepper, 2018).

### 3.1. Assumption Forms and Implications

The simplest bounded-variation assumptions place a priori bounds directly on long mean treatment outcomes, expressing expert judgement about the potential range in which they may take values. Formally, one may believe it credible to assume that

$$(4) \quad a(t, \xi) \;\leq\; E[y(t)|x = \xi] \;\leq\; b(t, \xi),$$

where $a(t, \xi)$ and $b(t, \xi)$ are specified constants.

One may believe it credible to bound the degree to which mean outcomes vary across persons with different covariates. One may express this as a bound on the absolute or relative difference between mean outcomes. Such bounds have the form

(5) $c(t, \xi, \xi') \leq E[y(t)|x = \xi] - E[y(t)|x = \xi'] \leq d(t, \xi, \xi')$,

(6) $e(t, \xi, \xi') \leq E[y(t)|x = \xi]/E[y(t)|x = \xi'] \leq f(t, \xi, \xi')$,

where $(\xi, \xi')$ are distinct covariate values and $[c(t, \xi, \xi'), d(t, \xi, \xi'), e(t, \xi, \xi'), f(t, \xi, \xi')]$ are specified constants.

Or one may believe it credible to bound the degree to which mean outcomes vary across treatments, expressing this as a bound on ATEs or on relative risks across treatments. Such bounds have the form

(7) $g(t, t', \xi) \leq E[y(t)|x = \xi] - E[y(t')|x = \xi] \leq h(t, t', \xi)$,

(8) $i(t, t', \xi) \leq E[y(t)|x = \xi]/E[y(t')|x = \xi] \leq j(t, t', \xi)$.

Here $(t, t')$ are distinct treatments and $[g(t, t', \xi), h(t, t', \xi), i(t, t', \xi), j(t, t', \xi)]$ are specified constants.

Bounded-variation assumptions of forms (4)−(8) are not mutually exclusive. One may believe it credible to assert a set of such assumptions, depending on the context. An interesting consequence of equations (2) is that assumptions constraining long mean outcomes at specified covariate values $\xi$ or $(\xi, \xi')$ help to identify long mean outcomes at other covariate values as well.

Adding assumptions of forms (4)−(6) to the information available in (1)−(3) does not complicate the problem of computing identification regions for long mean outcomes. It is straightforward to add these inequalities. We have found that our computational approach works well in practice when these inequalities are imposed. In contrast, adding assumptions of forms (7)−(8) complicates computation substantially. ne must now jointly apply equations (2) to both treatments t and t′. This increases the dimensionality and complexity of the problem of searching for sharp lower and upper bounds.

3.2. Further Illustrative Application

We add bounded-variation assumptions to the illustration in Section 2.3. Specifically, we add symmetric bounds on ATEs of form (5) when $\xi$ and $\xi'$ differ from one another in only one component. Formally, we choose a constant $b > 0$ and add the assumption that $-b \leq E[y(e)|x = \xi] - E[y(p)|x = \xi'] \leq b$ whenever $\xi$ and $\xi'$ differ from one another in only one of the three components $(x_1, x_2, x_3)$.

We report here computations when b = 0.05. Tables 2A−2B present the findings.

Table 2A: Sharp Bounds on Long Mean Outcomes with Treatment by Empagliflozin when b = 0.05

| **x label** | **P[y(e) = 1\|x] lower bound** | **P[y(e) = 1\|x] upper bound** |
|---|---|---|
| YML | 0.060 | 0.133 |
| OML | 0.063 | 0.159 |
| YFL | 0.033 | 0.144 |
| OFL | 0.047 | 0.153 |
| YMH | 0.056 | 0.160 |
| OMH | 0.067 | 0.178 |
| YFH | 0.020 | 0.155 |
| OFH | 0.052 | 0.191 |

Table 2B: Sharp Bounds on Long Mean Outcomes with Treatment by Placebo when b = 0.05

| **x label** | **P[y(p) = 1\|x] lower bound** | **P[y(p) = 1\|x] upper bound** |
|---|---|---|
| YML | 0.103 | 0.136 |
| OML | 0.152 | 0.173 |
| YFL | 0.056 | 0.099 |
| OFL | 0.105 | 0.142 |
| YMH | 0.055 | 0.094 |
| OMH | 0.105 | 0.142 |
| YFH | 0.105 | 0.142 |
| OFH | 0.134 | 0.192 |

We found that weaker assumptions using the larger constants b = 0.1 and b = 0.2 yielded informative upper bounds but mainly uninformative lower bounds on long mean outcomes.

In presenting these findings, we do not assert that medical decision makers considering treatment of diabetes should believe it credible to assert the version of assumption (5) posed here. Our direct objective is simply to illustrate identification using bounded-variation assumptions. Should one think it reasonable to make the specific assumption used to produce Tables 2A-2B, we point out that we obtain fairly narrow bounds on long mean outcomes, with widths always less than 0.135 for $t = e$ and less than 0.058 for $t = p$. Nevertheless, these bounds are not sufficiently informative to conclude that one treatment outperforms the other in terms of long ATE or relative risk.

## 4. Measuring Sampling Imprecision

The prevalent approach to sample inference on partially-identified parameters has been to compute confidence sets having desirable asymptotic properties. However, it appears that the methods developed to date, reviewed in Canay and Shaikh (2017) and Molinari (2020), cannot be applied in the present setting.

The identification region for $\{E[y[t)|x], P(x)]\}$ is a set-valued function of the vector of short quantities $\{E[y(t)|x_k = \xi_k], P(x_k = \xi_k), \xi_k \in \{0, 1\}, k = 1, \ldots, K\}$. These are estimated in trial findings by sample averages $\{E_N[y(t)|x_k = \xi_k], P_N(x_k = \xi_k), \xi_k \in \{0, 1\}, k = 1, \ldots, K\}$, where N is the number of subjects who receive treatment t. If subjects are randomly drawn from the population and randomly assigned to treatments, each component of this vector of averages has a limiting normal distribution centered on the corresponding population quantity. However, knowledge of these univariate limiting distributions does not suffice to form an asymptotically valid confidence set for $\{E[y[t)|x], P(x)]\}$ or features thereof. To accomplish this, one needs to know the limiting multivariate normal distribution of the vector of sample averages. The difficulty is that the reported trial findings do not provide an empirical basis to estimate the covariances between the sample estimates.

In the absence of an approach to compute confidence sets with provable good properties, an alternative way to examine sampling imprecision is to specify a suitable collection of hypothetical multivariate distributions for [y(·), x]. Given such a distribution, one may use Monte Carlo simulation to draw repeated pseudo-samples of observable data, compute repeated pseudo-estimates of identification regions for long mean outcomes, and examine their sampling variation across Monte Carlo repetitions. In principle, analysis of this sort can yield useful information. In practice, it challenging to implement given the need to repeatedly compute estimates of identification regions, a formidable task in the present setting.

## 5. Discussion

We have studied identification of long mean treatment outcomes using trial findings on outcomes in binary subgroups. Reporting outcomes in binary subgroups is the canonical form of a broad class of inferential problems that may arise in practice. Covariates may be multi-valued rather than binary. Articles could report mean treatment outcomes conditional on more than single covariates but less than all observed covariates. Our abstract identification analysis can be extended to these settings.

We have found that identification regions have simple abstract forms. Effective computation poses challenges, but it is tractable. Our computations indicate that the identification problem is severe when using reported trial findings alone. Bounded-variation assumptions have identifying power. However, one should keep in mind the Law of Diminishing Credibility (Manski, 2003): The credibility of inference decreases with the strength of the assumptions maintained.

Our approach relies on standardized reporting of credible subgroup analyses. Although multiple guidances have formulated these standards, the reporting of subgroup analyses in clinical trials remains problematic even in top journals (Gabler *et al.*, 2016). Medical journals that publish randomized trials should take responsibility in improving the reporting.

Sharing individual-level patient data could be a more beneficial solution for personalized treatment. If individual-level patient data becomes available, users should recognize that refinement of subgroup analysis to condition on multiple patient covariates diminishes sample size, reducing the sampling precision of estimates. Concern with imprecision is legitimate. However, the longstanding focus of medical research on conventional measures of statistical significance of estimated treatment effects does not provide a sound reason to avoid refinement of subgroup analysis. The traditional pre-occupation with statistical significance stems from the use of classical hypothesis tests to compare "standard care" treatments with innovations. Analysis by Manski and Tetenov (2019, 2021) shows that sample sizes which are too small to yield statistically significant estimates of treatment effects can nevertheless usefully inform treatment choice.

The main impediment to access to individual-level data in medical research has been institutional. Sharing individual-level trial data has been rare (Iqbal *et al.*, 2016). Efforts have been made to encourage data sharing (Taichman *et al.* 2016), but with only limited success (Taichman *et al.* 2017). The focus of such efforts has been to encourage data sharing among researchers, particularly to enable the replication of published findings. Replication is a worthy objective, but it only serves to corroborate findings published with current reporting conventions. We see a much more ambitious reason to share individual-level data, this being to enable clinicians and guideline developers to draw conclusions that would improve personalized treatment choice.

The Data and Code Availability Policy of the American Economic Association (2022b**)** states: "It is the policy of the American Economic Association to publish papers only if the data and code used in the analysis are clearly and precisely documented and access to the data and code is non-exclusive to the authors." We believe that medicine could usefully emulate economics in its position on data sharing. Nevertheless, before data sharing in the discipline of medicine becomes widely available, our novel approach of partial identification with bounded variation assumptions could facilitate clinicians' decision-making in practice.

Appendix 1: Proof that Identification Region is an Interval When Outcomes are Binary

To show that the identification region is the entire interval, define $w(t, \xi) \equiv P[y(t) = 1, x = \xi]$ and $P(\xi) \equiv P(x = \xi)$. Assumption (1b) states that $P(\xi) > 0$. Hence, $E[y(t)|x = \xi) = w(t, \xi)/P(\xi)$.

We first show that the set of feasible values of $[w(t, \xi), P(\xi)]$ is convex. To see this, let $[w_0(t, \xi), P_0(\xi)]$ and $[w_1(t, \xi), P_1(\xi)]$ be any two vectors that are known to be feasible. The set of feasible $[w(t, \xi), P(\xi)]$ is convex if

$$[w_c(t, \xi), P_c(\xi)] \equiv c \cdot [w_1(t, \xi), P_1(\xi)] + (1 - c) \cdot [w_0(t, \xi), P_0(\xi)]$$

is feasible whenever $0 \le c \le 1$. Inspection of (1)−(3) shows immediately that $c \cdot P_1(\xi) + (1 - c) \cdot P_0(\xi)$ is a feasible multivariate covariate probability. It is a bit less obvious whether $c \cdot w_1(t, \xi) + (1 - c) \cdot w_0(t, \xi)$ is feasible. Given the definition of $w(t, \xi)$ as a joint probability, the feasibility condition is that

$$c \cdot w_1(t, \xi) + (1 - c) \cdot w_0(t, \xi) \le c \cdot P_1(\xi) + (1 - c) \cdot P_0(\xi) = P_c(\xi).$$

This holds, so $w_c(t, \xi)$ is feasible.

It remains to ask whether convexity of the set of feasible values of $[w(t, \xi), P(\xi)]$ implies convexity of the set of feasible values of $E[y(t)|x = \xi]$. The answer is positive. For $0 \le c \le 1$,

$$E_c[y(t)|x = \xi] = w_c(t, \xi)/P_c(\xi) = [c \cdot w_1(t, \xi) + (1 - c) \cdot w_0(t, \xi)]/[c \cdot P_1(\xi) + (1 - c) \cdot P_0(\xi)].$$

The denominator is positive by assumption. Hence, the expression on the right-hand side is continuous over $c \in [0,1]$. It follows, by the intermediate value theorem, that as c increases from 0 to 1, $E_c[y(t)|x = \xi]$ takes all values in the interval connecting $E_0[y(t)|x = \xi]$ and $E_1[y(t)|x = \xi]$.

Appendix 2: Computational Approach

Equations (2)−(3) are differentiable. It is reasonable to conjecture that standard nonlinear local or global solvers might be able to handle the computational problem of finding the extrema of $E[y(t)|x = \xi]$. The issue is to find an approach that works adequately in practice.

Exploration revealed obstacles to the reliable use of a purely local or global solver. Consider the former. We do not have a robust way to run a nonlinear local solver from a single starting point and be confident that we have found the true extremum.

Consider the latter. The complete long vector of unknowns, $\{E[y(t)|x = \xi], P(x = \xi)\}\ \xi \in \{0, 1\}^K$, is the $2^{K+1}$-dimensional unit hypercube. Searching this space with a global solver, such as simulated annealing, quickly becomes intractable as K increases.

Given these considerations, we chose to use the *MATLAB Global Optimization Toolbox*, which provides a widely available and flexible set of solvers. Within the toolbox, we use the *GlobalSearch* solver, described in Matlab (2022). This hybrid local-global approach runs the *fmincon* local solver at multiple starting points in the search space. Thus, before choosing a single output value, *GlobalSearch* considers a set of possible local extrema (*fmincon* solutions from different starting points) and chooses the best of these. Importantly, *fmincon* accepts nonlinear constraints such as those in our equations (2).

We provide further detail on the solver here. We discuss search for the sharp lower bound on $E[y(t)|x = \xi\}$, search for the upper bound being analogous. *GlobalSearch* considers a local solution to be approximately feasible if its maximal violation of a specified constraint is smaller than a specified precision threshold. If the algorithm finds multiple such local solutions, it chooses the one yielding the most extreme value of the quantity being optimized.

It may be the case that the algorithm finds no approximately feasible local solutions. It then seeks to select among values that are substantially infeasible, in the sense of violating a constraint nontrivially. In such cases, *GlobalSearch* chooses the value with the lowest score, where the score is the sum of the objective function and the sum of all constraints times some constraint-weighting multiple. Matlab (2022) states: "*[t]he multiple for constraint violations is initially 1000. GlobalSearch updates the multiple during the run.*" No further discussion explains how this multiple is updated. Presumably it is increased if the solver is stuck on an infeasible value in order to find the *least infeasible* value in a certain local basin of attraction. Our analysis of the level of constraint violations of infeasible solutions seems to support this claim.

Thus, it appears that *GlobalSearch* only chooses an infeasible point if the constraint violation is relatively small, and the value of the objective function is much smaller than the best feasible alternative. Through this heuristic, *GlobalSearch* seeks to provide results which, although strictly infeasible, may still be judged useful and reasonable. The presumption seems to be that the locally smooth structure of the extremum problem gives a reasonable assurance that the 'real' solution is nearby to where the solver found an infeasible solution.